\documentclass[12pt]{article}
\pdfoutput=1
\usepackage{amssymb,amsmath}
\usepackage{epsfig}
\usepackage{epstopdf}
\usepackage{latexsym}
\usepackage{graphicx}
\usepackage{booktabs}
\usepackage{bbm}
\usepackage{enumitem}
\usepackage[numbers,compress]{natbib}

\usepackage[T1]{fontenc}

\usepackage[margin=20pt,small]{caption}
\usepackage{subcaption}

\usepackage[toc]{appendix}

\usepackage{color}
\usepackage{datetime}
\usepackage[
      colorlinks=false,
      linkcolor=darkblue,  
      urlcolor=blue,    
      filecolor=blue,     
      citecolor=red,
linktocpage=true,
      pdfstartview=FitV,
      bookmarksopen=true,
	  hidelinks
      ]{hyperref}

\ifpdf
\fi

\newcommand*{\boxedcolor}{red}
\makeatletter
\renewcommand{\boxed}[1]{\textcolor{\boxedcolor}{%
  \fbox{\normalcolor\m@th$\displaystyle#1$}}}
\makeatother

\definecolor{cardinal}{rgb}{0.6,0,0}
\definecolor{darkgreen}{rgb}{0,0.5,0}
\definecolor{golden}{rgb}{0.92, 0.7, 0}
\definecolor{midnight}{rgb}{0, 0, 0.5}
\definecolor{darkblue}{rgb}{0.2, 0, 0.8}

\def\im{{\rm i}} 

\def\sl{\frak{sl}}

\newcommand{\be}{\begin{equation}}
\newcommand{\ee}{\end{equation}}
\newcommand{\bea}{\begin{eqnarray}}
\newcommand{\eea}{\end{eqnarray}}

\begin{document}  

\begin{titlepage}
 
\medskip
\begin{center} 
{\Large \bf  Superconformal Blocks for SCFTs with Eight Supercharges}

\bigskip
\bigskip
\bigskip
\bigskip

{\bf Nikolay Bobev,${}^{(1)}$ Edoardo Lauria,${}^{(1)}$ and  Dalimil Maz\'a\v c,${}^{(2)}$  \\ }
\bigskip
\bigskip
\bigskip
\bigskip
${}^{(1)}$
Instituut voor Theoretische Fysica, KU Leuven \\
Celestijnenlaan 200D, B-3001 Leuven, Belgium
\vskip 8mm
${}^{(2)}$ Perimeter Institute for Theoretical Physics \\
31 Caroline Street North, ON N2L 2Y5, Canada\\
\bigskip
\texttt{nikolay.bobev@kuleuven.be,~edo.lauria@kuleuven.be, dalimil.mazac@gmail.com} \\
\end{center}

\bigskip
\bigskip

\begin{abstract}

\noindent  
\end{abstract}

\noindent 

We show how to treat the superconformal algebras with eight Poincar\'{e} supercharges in a unified manner for spacetime dimension $2 < d\leq 6$. This formalism is ideally suited for analyzing the quadratic Casimir operator of the superconformal algebra and its use in deriving superconformal blocks. We illustrate this by an explicit construction of the superconformal blocks, for any value of the spacetime dimension, for external protected scalar operators which are the lowest component of flavor current multiplets.

\end{titlepage}


\setcounter{tocdepth}{2}
\tableofcontents


\section{Introduction}

The conformal bootstrap programme has evolved from a general set of constraints that consistent CFTs should obey \cite{Polyakov:1974gs,Ferrara:1974ny,Ferrara:1974pt,Ferrara:1973yt,Ferrara:1971vh,Ferrara:1974nf,Ferrara:1973vz} to an increasingly important tool to obtain quantitative information about strongly interacting CFTs \cite{Rattazzi:2008pe,ElShowk:2012ht}.\footnote{See also \cite{Rychkov:2016iqz,Poland:2016chs,Simmons-Duffin:2016gjk} for a review and a more comprehensive list of references.} The basic idea is to implement the constraints imposed by unitarity and crossing symmetry on the four-point functions in the CFT, to constrain the spectrum of local operators and OPE coefficients of the theory. To this end, one expands a four-point function in terms of the conformal blocks, which are functions capturing the contribution of a given conformal family exchanged in a fixed OPE channel. It is thus clear that conformal blocks are an important technical ingredient for the success of the conformal bootstrap programme. Despite the fact that conformal blocks are kinematical quantities, i.e. their functional form is entirely determined by the conformal symmetry of the theory, the explicit construction of these functions is in general an involved technical problem. In their pioneering work, \cite{Dolan:2000ut,Dolan:2003hv}, Dolan and Osborn showed how to find the conformal blocks for external scalar operators in general spacetime dimensions. In fact, it turns out that in this analysis the dimension of spacetime, $d$, appears as a parameter in the conformal block and one can treat (at least formally) CFTs in non-integer dimensions. The explicit form of these scalar conformal blocks are known in two, four and six dimensions while for other values of $d$ one needs to resort to a series expansion, see for example \cite{Dolan:2003hv,Penedones:2015aga,Costa:2016xah,Dolan:2011dv,Hogervorst:2013sma}.

It is reasonable to expect that supersymmetry constrains further the space of consistent CFTs and it is thus natural to apply the conformal bootstrap methods to supersymmetric CFTs. This program has had a lot of success recently with a plethora of explicit quantitative analytical and numerical results for SCFTs in various dimensions, see for example \cite{Poland:2010wg,Bashkirov:2013vya,Beem:2013sza,Beem:2013qxa,Alday:2013opa,Fitzpatrick:2014oza,Berkooz:2014yda,Khandker:2014mpa,Beem:2014kka,Alday:2014qfa,Li:2014gpa,Chester:2015qca,Beem:2015aoa,Bissi:2015qoa,Liendo:2015ofa,Poland:2015mta,Lemos:2015awa,Lin:2015wcg,Li:2016chh,Lin:2016gcl,Li:2017ddj,Cornagliotto:2017dup}. A prerequisite for these bootstrap studies is the explicit construction of the so-called superconformal blocks, i.e. the analog of conformal blocks for SCFTs. The goal of our work is to address the construction of superconformal blocks for theories with eight Poincar\'e supercharges\footnote{The closure of the superconformal algebra implies that these theories also posses eight conformal supercharges.} in general spacetime dimension in the range $2<d\leq 6$. Our choice for the upper bound on $d$ can be attributed to the fact that there are no superconformal algebras in more than six dimensions \cite{Nahm:1977tg}. As explained in Section 2, the reason to restrict to theories in more than two dimensions is more technical and is related to the existence of a family of superconformal algebras with eight supercharges in two-dimensions.

The main motivation in constructing explicitly superconformal blocks for SCFTs with eight supercharges is to understand the dynamics of these theories using conformal bootstrap methods. SCFTs with eight supercharges posses a rich mathematical structure and have proven to be very useful theoretical laboratories for understanding conformal field theories and RG flows. Such theories arise naturally in string and M-theory through various brane and geometric constructions. In particular SCFTs with eight (or more) supercharges provide the only known examples of unitary interacting CFTs in more than four spacetime dimensions, see for example \cite{Seiberg:1996bd,Seiberg:1996qx}. In addition, SCFTs with eight supercharges in $d=3,5,6$ are isolated \cite{Cordova:2016xhm,Buican:2016hpb} (see also \cite{Louis:2015mka}), i.e. they do not posses exactly marginal supersymmetric deformations. This fact makes these SCFTs particularly amenable to analysis using algebraic techniques like the conformal bootstrap. 

SCFTs with eight supercharges necessarily posses at least an $SU(2)$ R-symmetry group. In addition to that, almost all known examples of these theories have a continuous flavor symmetry group.\footnote{We call all global symmetries that commute with the supercharges of the SCFT flavor symmetries.} The conserved current associated to this flavor symmetry belongs to a short superconformal multiplet, the lowest component of which is a real scalar operator transforming in the spin-1 representation of $SU(2)_R$ and in the adjoint representation of the flavor group, see for example \cite{Cordova:2016emh,Buican:2016hpb}. In a slight abuse of notation, we will refer to these scalar operators as ``moment map'' operators. The four-point function of these operators is the main object of interest in this paper. In particular, we show how to expand this four-point function into superconformal blocks, which we explicitly compute in any spacetime dimension in the range $2<d\leq 6$. To achieve this, we need a unified language to discuss superconformal algebras with eight supercharges and their representations. A convenient way to approach this is to start with the $(1,0)$ superconformal algebra in six dimensions, which has an $SU(2)_R$ R-symmetry, and then obtain the lower-dimensional superconformal algebras as a formal dimensional reduction. One then finds the following familiar list of R-symmetry groups in integer dimensions. 
\begin{equation}
\label{eq:Rsymm}
\begin{array}{lll}
d=6&\quad{\color{red}SU(2)_R}&\quad \mathcal{N}=(1,0)\\
d=5&\quad {\color{red}SU(2)_R}&\quad\mathcal{N}=1\\
d=4&\quad {\color{red}SU(2)_R} \times  U(1)&\quad\mathcal{N}=2\\
d=3&\quad {\color{red}SU(2)_R} \times SU(2)&\quad\mathcal{N}=4
\end{array}
\end{equation}
The extra factors in the R-symmetry group in $d=4$ and $d=3$ can be thought of as arising from the rotation group in the ``transverse'' 2 and 3 dimensions respectively.\footnote{In $d=2$ the small superconformal algebra has an $SO(4)$ R-symmetry which can be fully accounted for by the rotation group in the $(3,4,5,6)$ directions. The ``universal'' $SU(2)_{R}$ is thus not present in $d=2$.} We can use this pattern as a suggestive hint and formulate, at least formally, the superconformal algebra with eight supercharges in any value of the spacetime dimension. This approach is similar to the one employed in \cite{Bobev:2015jxa} for superconformal algebras with four supercharges. Using this formal construction, we can easily study the quadratic Casimir of the superconformal algebra for any value of $d$. This operator is of particular importance for superconformal blocks since under certain conditions, these are eigenfunctions of the quadratic superconformal Casimir. For the four-point function of moment map operators, we are able to exploit this fact and derive differential equations for the corresponding superconformal blocks and demonstrate how to solve them explicitly.

This method for constructing superconformal blocks based on the quadratic superconformal Casimir operator follows closely the approach employed in \cite{Bobev:2015jxa,Bobev:2015vsa}. We want to stress that this is different from an explicit analysis of supersymmetric Ward identities using superspace or other methods \cite{Dolan:2001tt,Dolan:2004mu}. The results and methods of \cite{Dolan:2001tt,Dolan:2004mu} are the ones usually employed in the literature on superconformal blocks for three-dimensional $\mathcal{N}=4$ \cite{Chester:2014mea,Liendo:2016ymz}, $\mathcal{N}=6$ \cite{Liendo:2015cgi} and $\mathcal{N}=8$ \cite{Chester:2014fya} as well as four-dimensional\footnote{See also \cite{Doobary:2015gia} for another method to derive superconformal blocks for four-dimensional SCFTs.} $\mathcal{N}=2$ \cite{Beem:2014zpa}, $\mathcal{N}=3$ \cite{Lemos:2016xke}  and $\mathcal{N}=4$ CFTs \cite{Beem:2013qxa,Beem:2016wfs} SCFTs. Our approach can be viewed as a supersymmetric extension of the work of Dolan-Osborn who employed the fact that conformal blocks are eigenfunctions of the quadratic Casimir operator of the conformal algebra in non-supersymmetric CFTs \cite{Dolan:2003hv}.

We continue our story in the next section with a discussion on superconformal algebras with eight supercharges in general spacetime dimensions. This sets the stage for a discussion of the quadratic superconformal Casimir operator in Section \ref{sec:Casimireqn}. In Section \ref{sec:momentmap}, we use the quadratic Casimir operator to derive and solve differential equations in general spacetime dimensions obeyed by superconformal blocks for a particular class of external protected scalar operators. In Section \ref{sec:discussion}, we conclude with some comments and list a number of possible avenues for further developments. In Appendix \ref{sec:Freehyper}, we show how the four-point function of the moment map operators in the theory of a free hypermultiplet is decomposed explicitly in terms of our superconformal blocks.

\textit{Note added:} During the final stages of writing this manuscript, we became aware of the recent work \cite{Chang:2017xmr} which has some overlap with our results. In particular, the authors of \cite{Chang:2017xmr} derive superconformal blocks for the four-point function of scalar moment map operators using a method based on the results of \cite{Dolan:2004mu}. Our results agree with the ones in the third version of \cite{Chang:2017xmr} on the arXiv.\footnote{There were some typographical errors in the expressions for the short multiplet blocks in the first two versions of \cite{Chang:2017xmr}. }

\section{Superconformal algebras with eight supercharges}
\label{sec:algebra}
\subsection{Set-up}
To study superconformal algebras with eight supercharges for general values of the spacetime dimensions $d$, we follow the approach outlined in \cite{Bobev:2015jxa}, where a similar problem was addressed for superconformal algebras with four supercharges. We would like to stress that superconformal algebras are well-defined only in integer dimensions. Therefore, many of the formulae below should be considered as a collection of formal manipulations which reduce to the well-known superconformal algebras when $d$ is an integer.

The main idea is to start from the $d=6$ superconformal algebra with $(1,0)$ supersymmetry and obtain the algebras for smaller values of $d$ by a formal dimensional reduction. We work in the Euclidean signature and impose reality conditions consistent with unitarity in the Lorentzian signature as usual. Our notation is such that Latin indices run over the unreduced spacetime directions $i=1,\ldots,d$, while with hatted indices we denote the reduced directions $\hat i= d+1,\ldots,6$. The bosonic generators of the superconformal algebra include the momenta $P_i$, special conformal $K_i$ and dilation $D$ generators. In addition, we have the rotations in the unreduced dimensions $M_{ij}$ as well as the rotations in the reduced dimensions $M_{\hat i \hat j}$. As emphasized in \cite{Bobev:2015jxa}, it is important to formally keep the reduced rotations $M_{\hat i \hat j}$ for any value of $d$, although there are no such generators for integer values of $d>4$. The explicit bosonic commutation relations are
\begin{equation}
\begin{aligned} 
 \lbrack M_{ij},M_{kl}] &= -\im (\delta_{il}M_{jk} + \delta_{jk}M_{il} - \delta_{ik}M_{jl} - \delta_{jl}M_{ik})\,,\\
 \lbrack M_{\hat i \hat j},M_{\hat k \hat l}] &= -\im (\delta_{\hat i \hat l}M_{\hat j \hat k} + \delta_{\hat j \hat k}M_{\hat i \hat l} - \delta_{\hat i \hat k}M_{\hat j \hat l} - \delta_{\hat j \hat l}M_{\hat i \hat k})\,,\\
 [M_{ij},P_k] &= -\im(\delta_{jk}P_i - \delta_{ik}P_j)\,,\\
 [M_{ij},K_k] &= -\im(\delta_{jk}K_i - \delta_{ik}K_j)\,,\\ 
 [D,P_i] &= -\im P_i\,,\\
 [D,K_i] &= \im K_i\,,\\
 [P_i,K_j] &= -2 \im(\delta_{ij}D + M_{ij})\,,
\end{aligned}
\label{bosonicCA}
\end{equation}
with all other commutators vanishing. In addition to the generators of the conformal algebra, there is also the omnipresent $SU(2)_R$ symmetry, denoted in red in \eqref{eq:Rsymm}, whose generators are
\begin{equation}
R^{a}_{\phantom{a}b} = (\sigma_{A})^{a}_{\phantom{a}b}R_A\,,
\end{equation}
where $(\sigma_{A})^{a}_{\phantom{a}b}$ are the usual Pauli matrices. They commute with all conformal generators and obey the following algebra
\begin{equation}
[R_A,R_B] = \im\varepsilon_{ABC}R_C\,.
\end{equation}
The extra factors in the R-symmetry algebra for $d\leq4$ in \eqref{eq:Rsymm} can be thought of as arising from the rotations in the reduced dimensions generated by $M_{\hat i \hat j}$.

We adopt the following Hermitian conjugation rules
\begin{equation}
D^{\dagger} = -D\;, \qquad R_A^{\dagger} = R_A\;, \qquad M_{ij}^{\dagger} = M_{ij}\;, \qquad M_{\hat i \hat j}^{\dagger} = M_{\hat i\hat j}\;, \qquad P_{i}^{\dagger} = K_i\;.
\label{eq:conjbos}
\end{equation}
Note that in our conventions, the action of the dilation generator $D$ on an operator $\mathcal{O}$ is $[D, \mathcal{O}] = -\im\Delta \mathcal{O}$, where $\Delta$ is the conformal dimension of $\mathcal{O}$.

In addition to these bosonic operators, the superconformal algebra contains also eight Poincar\'e supercharges, $Q$, as well as eight conformal supercharges $S$. To describe the commutation relations obeyed by these generators, we momentarily focus on $d=6$. The Poincar\'e supercharges transform as a doublet of the $SU(2)_R$ R-symmetry and as a Weyl spinor of the $SO(6)$ rotations. We denote these supercharges as $Q_{a\alpha}$, where $a$ and $\alpha$ are the $SU(2)_R$, and spinor indices respectively. The anti-commutator of these fermionic generators takes the form
\begin{equation}
\{Q_{a\alpha},Q_{b\beta}\} = \epsilon_{ab}\Gamma_{\alpha\beta}^i P_i\,,
\end{equation}
where $\Gamma_{\alpha\beta}^i$ are a set of antisymmetric matrices satisfying extra conditions as discussed later. The conformal supercharges transform in the conjugate Weyl representation, denoted with an upper index. We also make use of the following conjugation rule
\begin{equation}
S^{a\alpha} = (Q_{a\alpha})^\dagger\,.
\end{equation}
With these definitions, index contraction makes sense since both $a$ and $\alpha$  correspond to unitary representations. Using the notation $\tilde{\Gamma}_{i}^{\alpha\beta} = (\Gamma^i_{\beta\alpha})^*$, the anti-commutator of the $S$ generators takes the form
\begin{equation}
\{S^{a\alpha},S^{b\beta}\} = \epsilon^{ab}\tilde{\Gamma}^{\alpha\beta}_i K_i\,.
\end{equation}
The $QQK$ Jacobi identity determines
\begin{equation}
[K_i,Q_{a\alpha}] = \epsilon_{ab}\Gamma_{\alpha\beta}^iS^{b\beta}\,,
\end{equation}
and hence
\begin{equation}
[P_i,S^{a\alpha}] = -\epsilon^{ab}\tilde{\Gamma}^{\alpha\beta}_iQ_{b\beta}\,.
\end{equation}
It follows from the $PKQ$ Jacobi identity that the $\Gamma$ matrices should obey the following identity
\begin{equation}
\tilde{\Gamma}_i\Gamma_j + \tilde{\Gamma}_j\Gamma_i = 2\delta_{i j}\,,
\label{eq:clifford}
\end{equation}
This identity is of course obeyed if we choose $\Gamma_i$ to be the usual Weyl matrices. The action of the rotation generators on the supercharges can be written as
\begin{equation}
\begin{aligned}
\lbrack M_{ij},Q_{a\alpha}] &= (m_{ij})^{\beta}_{\phantom{a}\alpha}Q_{a\beta}\,,\\
\lbrack M_{ij},S^{a\alpha}] &= - (m_{ij})^{\alpha}_{\phantom{a}\beta}S^{a\beta}\,,
\end{aligned}\label{MQcomm}
\end{equation}
The $PKQ$ Jacobi identity then leads to the relation
\begin{equation}\label{Weyl_rel}
m_{ij} = -\frac{\im}{4}(\tilde{\Gamma}_i\Gamma_j - \tilde{\Gamma}_j\Gamma_i )\,.
\end{equation}
As pointed out earlier, the supercharges are in the  doublet representation of $SU(2)_R$ and thus obey the following relations
\begin{equation}
\begin{aligned}
\lbrack R_A,Q_{a\alpha}] &= \frac{1}{2}(\sigma_A)^{b}_{\phantom{a}a}Q_{b\alpha}\,,\\
\lbrack R_A,S^{a\alpha}] &= - \frac{1}{2}(\sigma_A)^{a}_{\phantom{a}b}S^{b\alpha}\,,
\end{aligned}
\end{equation}
Employing various Jacobi identities one can then determine the following anti-commutator between the Poincar\'e and the conformal supercharges
\begin{equation}\label{eq:SQ6d}
\{S^{a\alpha},Q_{b\beta}\} = \im \delta^{a}_{\phantom{a}b}\delta^{\alpha}_{\phantom{a}\beta} D - 4 \delta^{\alpha}_{\phantom{a}\beta}R^{a}_{\phantom{a}b}+\delta^{a}_{\phantom{a}b}(m_{ij})^{\alpha}_{\phantom{a}\beta}M_{ij}\,.
\end{equation}
%

\subsection{Lowering the dimension}

All commutation and anticommutation relations for the supercharges $S$ and $Q$ above are valid for $d=6$. Moreover, one can show that all Jacobi identities, except $SQQ$, are formally satisfied in the above algebra when we let the spacetime vector indices, $i,j,\ldots$, run from $1$ to any $d\leq 6$, using only the Clifford algebra \eqref{eq:clifford}. In addition, we will take  the relations in \eqref{eq:clifford}, \eqref{MQcomm}, and \eqref{Weyl_rel} to hold also for the hatted indices, $\hat{i},\hat{j},\ldots$ which label the reduced dimensions. This action ultimately defines the action of the extra R-symmetry factors in \eqref{eq:Rsymm} for integer values of $d$. 

The $SQQ$ Jacobi identity requires a more careful treatment. To obey it, one has to modify the $SQ$ anti-commutator relation in \eqref{eq:SQ6d} by making the coefficient of the $SU(2)_R$ R-symmetry dimension-dependent, and include the rotations in the reduced dimensions. The result is the following anti-commutation relation
\begin{equation}
\{S^{a\alpha},Q_{b\beta}\} = \im \delta^{a}_{\phantom{a}b}\delta^{\alpha}_{\phantom{a}\beta} D - (d-2) \delta^{\alpha}_{\phantom{a}\beta}R^{a}_{\phantom{a}b}+\delta^{a}_{\phantom{a}b}(m_{ij})^{\alpha}_{\phantom{a}\beta}M_{ij}-
\delta^{a}_{\phantom{a}b}(m_{\hat i\hat j})^{\alpha}_{\phantom{a}\beta}M_{\hat i\hat j}\,.
\label{eq:sqd}
\end{equation}
As explained in the beginning of this section, one has to take the unhatted spacetime vector indices, $i,j$, to run from $1$ to $d$, and the hatted ones, $\hat{i},\hat{j}$,  from $d+1$ to 6. Note the negative sign in front of the term involving $M_{\hat i\hat j}$ on the right hand side of \eqref{eq:sqd}. This ensures the correct action of the extra R-symmetry factors in \eqref{eq:Rsymm}. It can be checked that this $d$-dependent modification of the anti-commutator in \eqref{eq:SQ6d} does not spoil any of the other Jacobi identities. To obey the $SQQ$ Jacobi identity, the Weyl matrices have to obey the following two quartic relations
\begin{equation}
(m_{ij})^{\alpha}_{\phantom{a}\beta}(m_{ij})^{\gamma}_{\phantom{a}\delta} -
(m_{\hat i \hat j})^{\alpha}_{\phantom{a}\beta} (m_{\hat i \hat j})^{\gamma}_{\phantom{a}\delta} + (\alpha\leftrightarrow\gamma) = \frac{d-3}{2}\delta^{\alpha}_{\phantom{a}\beta}\delta^{\gamma}_{\phantom{a}\delta}+ (\alpha\leftrightarrow\gamma)\,,
\end{equation}
\begin{equation}
\tilde{\Gamma}_i^{\alpha\gamma}\Gamma^i_{\beta\delta} = \frac{d-1}{2}\delta^{\alpha}_{\phantom{a}\delta}\delta^{\gamma}_{\phantom{a}\beta}-(d-2)\delta^{\alpha}_{\phantom{a}\beta}\delta^{\gamma}_{\phantom{a}\delta}+
(m_{ij})^{\alpha}_{\phantom{a}\delta}(m_{ij})^{\gamma}_{\phantom{a}\beta} -
(m_{\hat i \hat j})^{\alpha}_{\phantom{a}\delta} (m_{\hat i \hat j})^{\gamma}_{\phantom{a}\beta}\, .
\end{equation}
Remarkably, these relations can be checked to hold for any $d= 1,\ldots,6$. We do not know if they can be derived in a dimension-independent language but we will assume that they hold in the discussions below. Note however that the constant $d-2$ in \eqref{eq:sqd} in front of the original $SU(2)_R$ R-symmetry generators, $R^{a}_{\phantom{a}b}$, can be derived in a dimension-independent language by taking traces of the quartic relations, using the Clifford algebra, and the identities $\delta^{i}_{\phantom{a}i} = d$, $\delta^{\hat i}_{\phantom{a}\hat i} = 6 - d$. 

We can thus conclude that using the approach summarized above, we have a unified way to describe the superconformal algebras with eight Poincar\'e supercharges in any integer dimension. In addition, these formulae can be used for other purposes, e.g. for calculations involving the quadratic Casimir operator, for non-integer values of $d$. 

We would like to emphasize that for $d\leq2$ the discussion above is not  entirely valid since some generators decouple from the superconfomal algebra. In particular, the $SU(2)_R$ symmetry, denoted in red in \eqref{eq:Rsymm}, is not present and the $M_{\hat i\hat j}$ generators in the four reduced dimensions produce the $SO(4)$ R-symmetry of the two-dimensional ``small'' superconformal algebra.\footnote{We are not able to incorporate the $D(2,1;\alpha)$ family of ``large'' superconformal algebras in our formalism.} Due to this subtlety, we will restrict ourselves to the range $2< d\leq 6$ in the rest of this note.

\section{The superconformal Casimir equations}
\label{sec:Casimireqn}

\subsection{The four-point function of moment map operators}

In this note, we focus on the four-point function of the so-called \emph{moment map} operators. They are the superconformal primaries of the so-called $\mathcal{D}[0,1]$ multiplet. These operators are spacetime scalars of scaling dimension $\Delta = d-2$, transforming in the vector representation of $SU(2)_R$ and in the adjoint representation of the flavor group. The notation $\mathcal{D}[0,1]$ refers to $\ell=0$, $R=1$ of the lowest component. Upon acting on the superconformal primary with two $Q$ supercharges, one obtains a flavor current. 

We will denote the superconformal primaries by $\varphi^{A}$, where $A=1,2,3$ is the $SU(2)_R$ vector index. Since the flavor group commutes with the superconformal generators, it does not play a role in the construction of superconformal blocks and we will supress the adjoint flavor indices. The $\mathcal{D}[0,1]$ multiplet in general $d$ is the dimensional reduction of the $\mathcal{D}[0,1]$ multiplet in $d=6$. As a consequence of this fact, $\varphi^{A}$ are neutral under the $SO(6-d)$ R-symmetry coming from rotations in the reduced dimensions.

Conformal symmetry implies that the four-point function of moment map operators takes the following form
\be
\langle \varphi^A (x_1) \varphi^B (x_2) \varphi^C (x_3) \varphi^D (x_4) \rangle = 
\frac{1}{(|x_{12}||x_{34}|)^{2(d-2)}}
\mathcal{F}^{ABCD}(z,\bar{z})\,,
\label{eq:4ptF}
\ee 
where $z$ and $\bar{z}$ are defined by
\begin{equation}\label{crossrr}
z \bar{z}\equiv\frac{x_{12}^2 x_{34}^2}{x_{13}^2 x_{24}^2}\;, \qquad\qquad (1-z)(1-\bar{z})\equiv\frac{x_{14}^2 x_{23}^2}{x_{13}^2 x_{24}^2}\;.
\end{equation}

The $SU(2)_R$ symmetry ensures that the operators exchanged in the $s$-channel OPE must transform in either $R=0$, $R=1$, or $R=2$ representations of $SU(2)_R$. The function $\mathcal{F}^{ABCD}(z,\bar{z})$ can be decomposed accordingly as
\be
\mathcal{F}^{ABCD}(z,\bar{z}) = 
\sum\limits_{R=0}^2 Y_{R}^{ABCD}
\mathcal{F}^R(z,\bar{z})\,,
\label{eq:Fdecomposition}
\ee
where $Y_{R}^{ABCD}$ are the $SU(2)$ eigentensors, taking the following form
\begin{align}\label{Ys}
Y_0^{ABCD}\equiv &~\delta^{AB}\delta^{CD}\;,\nonumber\\
Y_1^{ABCD}\equiv &~ \delta^{AD}\delta^{BC}-\delta^{AC}\delta^{BD}\;,\\
Y_2^{ABCD}\equiv &~ 3\delta^{AC}\delta^{BD}+3\delta^{AD}\delta^{BC}-2\delta^{AB}\delta^{CD}\;.\nonumber
\end{align}
The $s$-channel OPE leads to the following decomposition of each $\mathcal{F}^R$
\be\label{eq:OPEF}
\mathcal{F}^R(z,\bar{z}) = \sum\limits_{\mathcal{P}} c_{\varphi\varphi\mathcal{P}}^2g_{\Delta_{\mathcal{P}},\ell_\mathcal{P}}(z,\bar{z})\,,
\ee
where the sum runs over conformal primary operators transforming as the symmetric traceless tensors of $SO(d)$, and $g_{\Delta,\ell}(z,\bar{z})$ are the corresponding conformal blocks. Note that all of the primaries appearing in this OPE expansion transform trivially under the $SO(6-d)$ R-symmetry since the same holds true for the moment maps.

Superconformal symmetry further relates the coefficients $c_{\varphi\varphi\mathcal{P}}^2$ in \eqref{eq:OPEF} of different conformal primaries from the same superconformal multiplet. This means that $\mathcal{F}^{ABCD}(z,\bar{z})$ can be expanded in terms of the \emph{superconformal blocks} $\mathcal{G}^{ABCD}_{\mathcal{O}}(z,\bar{z})$
\be
\mathcal{F}^{ABCD}(z,\bar{z}) = \sum\limits_{\mathcal{O}}
c_{\varphi\varphi\mathcal{O}}^2\mathcal{G}^{ABCD}_{\mathcal{O}}(z,\bar{z})\,,
\ee
where the sum runs over superconformal primaries. Each superconformal block is a sum over R-symmetry components as follows
\be
\mathcal{G}_{\mathcal{O}}^{ABCD}(z,\bar{z}) = 
\sum\limits_{R=0}^2 Y_{R}^{ABCD}
\mathcal{G}^R_{\mathcal{O}}(z,\bar{z})\,.
\ee

As we explain later, in our case the coefficient of each conformal primary is fixed in terms of the coefficient of its corresponding superconformal primary. This implies that the functions $\mathcal{G}^{ABCD}_{\mathcal{O}}(z,\bar{z})$ are fully fixed by the superconformal symmetry. We will be able to use the superconformal Casimir equation to find them in a closed form. To this end, let us first derive the form of the superconformal Casimir operator.

\subsection{The superconformal Casimir operator}

The quadratic superconformal Casimir operator, $C$, must be a linear combination of the quadratic Casimir, $C_b$, of the conformal subalgebra, the quadratic Casimir of the $SO(6-d)$ group of ``transverse'' rotations, $\frac{1}{2}\hat{M}_{ij}\hat{M}_{ij}$, and R-symmetry $R_A R_A$, as well as terms quadratic in the fermionic generators. The form of this operator is completely fixed by the requirement that $C$ commutes with all generators of the superconformal algebra. We find
\begin{align}
C=C_b+\frac{1}{2}[S^{a\alpha},Q_{a\alpha}]-(d-2)R_A R_A+\frac{1}{2}\hat{M}_{ij}\hat{M}_{ij}\;,
\label{eq:supercasimir}
\end{align}
where the quadratic Casimir operator of the conformal algebra is given by
\begin{equation}
C_b=\frac{1}{2}M_{ij}M_{ij}-D^2-\frac{1}{2}(P_i K_i+K_i P_i)\;,
\end{equation}
and as usual we have assumed summation over repeated indices.

Let us now consider a superconformal primary with dimension $\Delta$, transforming as a symmetric traceless tensor of spin $\ell$ under $M_{ij}$ (recall that by construction this primary operator is a singlet under $\hat{M}_{ij}$) and with $SU(2)_R$ charge $R$. Using the superconformal algebra, it is easy to check that
\begin{equation}\label{seigen}
[C,\mathcal{O}_{\Delta,\ell,R}]=\lambda_{c}\mathcal{O}_{\Delta,\ell,R}\;,
\end{equation}
where
\begin{equation}\label{lambdaC}
\lambda_{c}\equiv\lambda_{C_b}+4\Delta-(d-2)R(R+1)\;,
\end{equation}
and  
\begin{equation}\label{lambdaCb}
\lambda_{C_b}\equiv\Delta(\Delta-d)+\ell(\ell+d-2)\;,
\end{equation}
is the eigenvalue of $C_b$.

It was emphasized in \cite{Dolan:2003hv} that the conformal blocks, $g_{\Delta,\ell}$ in \eqref{eq:OPEF}, are eigenfunctions of the quadratic Casimir operator of the conformal algebra, $C_b$, with eigenvalue $\lambda_{C_b}$ given by \eqref{lambdaCb}. This fact was then used in \cite{Dolan:2003hv} to derive differential equations for the functions $g_{\Delta,\ell}(z,\bar{z})$. The same logic can be applied to the quadratic Casimir operator of the superconformal algebra in order to find differential equations for the superconformal blocks $\mathcal{G}^R_{\mathcal{O}}(z,\bar{z})$. This procedure was successfully implemented for theories with four supercharges in \cite{Bobev:2015jxa} and we will apply the same method for the case of eight supercharges below.

In order to arrive at differential equations satisfied by the superconformal blocks, we need to act with the Casimir operator, $C$, on the operators at positions $x_1$ and $x_2$ in the four-point function \eqref{eq:4ptF}. This action will in general mix different four-point functions. However, we can get decoupled differential equations by making special choices of the external $SU(2)_R$ indices. Let us first introduce the following convenient basis for the $SU(2)_R$ vector indices
\begin{align}\label{pmbasis}
\varphi^+\equiv\frac{1}{\sqrt{2}}(\varphi^1- \im \varphi^2)\;,& \qquad \varphi^-\equiv\frac{1}{\sqrt{2}}(\varphi^1+ \im \varphi^2)\;, \qquad \varphi^{0} \equiv \varphi^3\;.
\end{align}
Considering the action of the superconformal Casimir operator on the $\langle\varphi^+\varphi^+\varphi^-\varphi^-\rangle$ and $\langle\varphi^+\varphi^-\varphi^+\varphi^-\rangle$ correlators leads to two independent differential equations, discussed in the next subsections, which allow us to fix the superconformal blocks completely.

\subsection{The $\langle\varphi^+\varphi^+\varphi^-\varphi^-\rangle$  Casimir equation}
\label{chir_channell}

After specializing the external $SU(2)_R$ indices to $\langle\varphi^+\varphi^+\varphi^-\varphi^-\rangle$, we find from \eqref{eq:Fdecomposition} that only the $R=2$ component contributes
\be
\mathcal{F}^{++--}(z,\bar{z}) = 6\mathcal{F}^2(z,\bar{z})\,.
\ee

Let us first understand how the various terms in the superconformal Casimir \eqref{eq:supercasimir} act on the $\langle\varphi^+\varphi^+\varphi^-\varphi^-\rangle$ of the four-point function. The conformal Casimir operator $C_b$ acts as the usual non-supersymmetric differential operator, $\mathcal{D}_{\textrm{DO}}$, employed by Dolan and Osborn \cite{Dolan:2003hv}. The action of the second term in \eqref{eq:supercasimir}, containing the fermionic generators, can be simplified by the following equations
\be 
[S^{a\alpha},\varphi^A(0)] = 0\,,\quad [Q_{1\alpha},\varphi_{+}(x)] = [Q_{2\alpha},\varphi_{-}(x)] = 0\,.
\ee
The first identity above is a consequence of the fact that $\varphi^A$ is a superconformal primary, while the latter two are special cases of the BPS condition satisfied by $\varphi^A$. A short computation shows that $\frac{1}{2}[S^{a\alpha},Q_{a\alpha}]$ then acts as a scalar multiplication by $8\Delta_{\varphi}=8(d-2)$. The third term in \eqref{eq:supercasimir}, including the minus sign, acts as a scalar multiplication by $-(d-2)R_{\mathcal{P}}(R_{\mathcal{P}}+1)=-6(d-2)$ since $\mathcal{F}^2(z,\bar{z})$ only receives contributions from conformal primaries $\mathcal{P}$ with $R_{\mathcal{P}}=2$. The last term in \eqref{eq:supercasimir} gives zero.

When we restrict $\mathcal{F}^2(z,\bar{z})$ to the contributions coming from a fixed superconformal family, the Casimir must act by scalar multiplication by $\lambda_C$. Hence, we find the following differential equation for the $\mathcal{G}^2(z,\bar{z})$ component of the superconformal block
\begin{equation}
\left[\mathcal{D}_{\textrm{DO}}+2(d-2)\right]\mathcal{G}^2(z,\bar {z}) = \lambda_{C}\mathcal{G}^2(z,\bar {z})
\,.
\label{eq:sc1}
\end{equation}
The differential operator $\mathcal{D}_{\textrm{DO}}$ is the same as the one found in \cite{Dolan:2003hv}. 
\begin{equation}
\begin{aligned}
\mathcal{D}_{\textrm{DO}} \equiv & ~ 2z^2(1-z)\partial^2 + 2\bar{z}^2(1-\bar{z})\bar{\partial}^2  - 2(z^2\partial + \bar{z}^2\bar{\partial}) \\ &\qquad\qquad\qquad+ 2(d-2)\frac{z\bar{z}}{z-\bar{z}}\left[(1-z)\partial - (1-\bar{z})\bar{\partial}\right]\,.
\end{aligned}
\end{equation}

Since Equation \eqref{eq:sc1} takes the form of the usual differential equation satisfied by non-supersymmetric conformal blocks, we can conclude that any nonzero solution is a single conformal block corresponding to a conformal primary $\mathcal{P}$ with R-charge $R_{\mathcal{P}}=2$, conformal dimension $\Delta_\mathcal{P}$ and spin $\ell_\mathcal{P}$.

Since $\mathcal{P}$ must be a symmetric traceless tensor, \eqref{eq:sc1} imposes the following constraint between $\Delta$, $\ell$, $R$, $\Delta_\mathcal{P}$ and $\ell_{\mathcal{P}}$
\begin{equation}\label{constr_eqig}
\Delta_{\mathcal{P}}(\Delta_{\mathcal{P}}-d)+\ell_{\mathcal{P}}(\ell_{\mathcal{P}}+d-2)+2(d-2)=
\Delta(\Delta-d+4)+\ell(\ell+d-2)-(d-2)R(R+1).
\end{equation}
We proceed in the next subsection, where we obtain a differential equation involving also the $\mathcal{G}^0$ and $\mathcal{G}^1$ components of the superconformal blocks.

\subsection{The $\langle\varphi^+\varphi^-\varphi^+\varphi^-\rangle$  Casimir equation}

Let us now turn our attention to the $\langle\varphi^+\varphi^-\varphi^+\varphi^-\rangle$ component of the superconformal Casimir equation. First, it follows from \eqref{eq:Fdecomposition} that
\be
\mathcal{F}^{+-+-}(z,\bar{z}) = \mathcal{F}^0(z,\bar{z})+\mathcal{F}^1(z,\bar{z})+\mathcal{F}^2(z,\bar{z})\,.
\ee
Next, we have to analyze how the various terms in the superconformal Casimir \eqref{eq:supercasimir} act in this case. As previously, the first term of \eqref{eq:supercasimir} acts as the Dolan-Osborn differential operator $\mathcal{D}_{\textrm{DO}}$. With some work, following the same logic as detailed in Section 3.2 of \cite{Bobev:2015jxa}, the second term in \eqref{eq:supercasimir} can be written as a differential operator
\begin{equation}
\frac{1}{2}[S^{a\alpha},Q_{a\alpha}]~~\mapsto~~ \mathcal{D}_{\textrm{SQ}}\equiv4\left[z(1-z)\partial+\bar{z}(1-\bar{z})\bar{\partial}\right]\,.
\end{equation}
The third term in \eqref{eq:supercasimir} becomes a multiplication by the $SU(2)_R$ Casimir eigenvalue times $-(d-2)$, and the last term vanishes. Hence, the final equation obeyed by the superconformal blocks in this channel takes the form
\begin{align}
\left(\mathcal{D}_{\textrm{DO}}+\mathcal{D}_{\textrm{SQ}}\right)\sum_{R=0}^2 \mathcal{G}^{R}(z,\bar z)
=(d-2)[2\mathcal{G}^1(z,\bar z)+6\mathcal{G}^2(z,\bar z)]+ \lambda_{C}\sum_{R=0}^2 \mathcal{G}^{R}(z,\bar z)\,,
\label{eq:sc2}
\end{align}
with $\lambda_C$ defined in \eqref{lambdaC}.

One may wonder about the meaning of the remaining Casimir equation involving only $A=\pm$ components, namely the equation for the $\langle\varphi^+\varphi^-\varphi^-\varphi^+\rangle$ correlator. It can be obtained from the equation for $\langle\varphi^+\varphi^-\varphi^+\varphi^-\rangle$ by simultaneous application of $(-1)^R$ and the swap of coordinates $x_1\leftrightarrow x_2$. Consequently, a solution of equation \eqref{eq:sc2} will solve the equation following from the $\langle\varphi^+\varphi^-\varphi^-\varphi^+\rangle$ correlator if the expansion only involves conformal primaries $\mathcal{P}$ with uniform $(-1)^{R_{\mathcal{P}}+\ell_{\mathcal{P}}}$, i.e.
\be
(-1)^{R_{\mathcal{P}}+\ell_{\mathcal{P}}} = (-1)^{R_{\mathcal{O}}+\ell_{\mathcal{O}}}\,,
\label{parity_cons}
\ee
where $\mathcal{O}$ is the superconformal primary. In fact, this ``parity'' constraint \eqref{parity_cons} will be fundamental to fix the parameters in our superconformal blocks. Note that the same ``parity'' constraint for $d=4$ was discussed in Section 3.1.1 of \cite{Beem:2014zpa}.

\section{Superconformal blocks for moment map operators}
\label{sec:momentmap}

\subsection{Unitary multiplets in $d=3,4,5,6$}
\label{unit-cond}

In order to derive the superconformal blocks for moment map four-point functions, it is instructive to collect some well-known facts about the structure of unitary multiplets of the superconformal algebras with eight supercharges for $2< d\leq 6$. The unitary representations of superconformal algebras with extended supersymmetry have been studied by many authors beginning with the pioneering work in \cite{Dobrev:1985qv,Dobrev:1985vh,Dobrev:1985qz}. For the results summarized below, we have also made use of the more recent work in \cite{Minwalla:1997ka,Dolan:2001tt,Bhattacharya:2008zy,Cordova:2016emh,Cordova:2016xhm,Buican:2016hpb}. A cursory look at these references makes it clear that the structure of superconformal multiplets of SCFTs with eight supercharges depends heavily on the dimension of spacetime and on the R-symmetry groups summarized in \eqref{eq:Rsymm}. Thus one may worry that our attempt to derive the superconformal blocks in a dimension-independent way is bound to fail. Fortunately, as we summarize below, there is a way around this apparent impasse.

The key observation is that the structure of unitary superconformal multiplets that can in principle appear in the OPE of two moment map operators is fairly uniform across spacetime dimensions. In particular, it was shown in \cite{Dolan:2001tt,Dolan:2004mu,Beem:2014zpa} that whenever a conformal primary appears in the OPE of two moment map operators, then also its corresponding superconformal primary appears. It follows that the superconformal primary is a symmetric traceless tensor of $SO(d)$. Moreover, the moment maps are neutral under the $SO(6-d)$ R-symmetry coming from rotations in the reduced dimensions, and so any operator appearing in their OPE is also neutral under these transformations. The list of unitary superconformal multiplets satisfying the above properties follows. With $\Delta$, $\ell$ and $R$ we denote the scaling dimension, spin and R-charge of the superconformal primary respectively.

\paragraph{$4\leq d\leq 6$}

\be
\begin{aligned}
\mathcal{L}[\Delta,\ell,R],& \qquad \Delta > (d-2)R+\ell+2d-6\,, \\
\mathcal{A}[\ell,R],& \qquad \Delta=(d-2)R+\ell+2d-6\,,\\
\mathcal{B}[\ell,R],&\qquad \Delta= (d-2)R+\ell+d-2\,,\\
\mathcal{C}[0,R],&\qquad \Delta= 4R+2\qquad\qquad (d=6)\,,\\
\mathcal{D}[0,R],&\qquad \Delta= (d-2)R\,.
\label{5dshort} 
\end{aligned}
\ee

\paragraph{$2<d\leq 4$}
\be
\begin{aligned}
\mathcal{L}[\Delta,\ell,R],& \qquad \Delta > (d-2)R+\ell+d-2\,, \\
\mathcal{B}[\ell,R],& \qquad \Delta= (d-2)R+\ell+d-2\,, \\
\mathcal{D}[0,R],&\qquad\Delta= (d-2)R.
\label{3dshort} 
\end{aligned}
\ee

The first lines in \eqref{5dshort} and \eqref{3dshort} correspond to long unitary multiplets. The second lines correspond to the short multiplet that emerges when the long multiplet reaches the unitarity bound. We call these \emph{regular} short multiplets. The remaining lines correspond to \emph{isolated} short multiplets. Note that the $\mathcal{B}$-type multiplet is isolated for $d>4$ but becomes regular in $d\leq 4$. For $R=\ell=0$, this multiplet contains the R-symmetry current and the energy-momentum tensor.

A few comments are in order. The expressions in \eqref{5dshort} and \eqref{3dshort} have been derived rigorously in the respective integer dimensions but we have written them in a suggestive way such that the dimension, $d$, appears as a parameter.  One should note that for $d=6$, there are the somewhat special $\mathcal{C}[0,R]$ multiplets which are due to the presence of self-dual tensor in six dimensions. In particular, $\mathcal{C}[0,0]$ is the $(1,0)$ free tensor multiplet. This multiplet and many others on the above list are in fact ruled out from appearing in the moment map OPE by our Casimir equation, as explained in the following section.

\subsection{The general logic of our derivation}

Crucially, not all multiplets listed in the previous subsection actually appear in the OPE of two moment map operators. For example, we can clearly restrict the R-charge to $R=0,1,2$. In addition, superconformal Ward identities further restrict the allowed set. In fact, our superconformal Casimir equations are powerful enough to sidestep the use of superconformal Ward identities. Indeed, the equations in \eqref{eq:sc1} and \eqref{eq:sc2} admit nonzero solutions only for the multiplets allowed by the Ward identities. Moreover, in the cases when a solution exists, it is unique and thus equal to the sought superconformal block.

Let us spell out our procedure in more detail. To determine a superconformal block means to find the functions $\mathcal{G}^{0}$, $\mathcal{G}^{1}$ and $\mathcal{G}^{2}$ for each allowed superconformal family. Each $\mathcal{G}^{R}$ is a finite linear combination of ordinary conformal blocks
\begin{equation}\label{ordblocks}
\mathcal{G}^{R}_{\mathcal{O}}(z,\bar z)=\sum_{n,m \in \mathbb{Z}} f_{n,m}^R g_{\Delta_{\mathcal{O}}+n,\ell_{\mathcal{O}}+m}(z,\bar z)\;,
\end{equation}
where $\mathcal{O}$ is the superconformal primary. Each element of the list of unitary multiplets with a symmetric traceless superconformal primary, presented in Section \ref{unit-cond}, provides an Ansatz for the set of conformal primaries that appear on the RHS of \eqref{ordblocks}. We can then apply the superconformal Casimir equations in \eqref{eq:sc1} and \eqref{eq:sc2} and fix the undetermined coefficients $f_{n,m}$. More specifically, we use the following representation of ordinary conformal blocks as a power series in $s=\sqrt{z \bar{z}}$, discussed in \cite{Hogervorst:2013sma} (see in particular Equation (2.24) of  \cite{Hogervorst:2013sma}).
 \begin{align}\label{confbb}
 g_{\Delta,\ell}(z,\bar{z})=\sum_{n=0}^\infty h_{n}(\xi)s^{\Delta + n}\,,
 \end{align}
where $\xi = (z+\bar{z})/2|z|$ and $h_{n}(\xi)$ can be determined recursively, starting from the initial condition
\be
\label{eq:gegenbauer}
h_0(\xi) = \frac{\ell !}{(2\nu)_{\ell}} C_{\ell}^{\nu} (\xi)\,,\qquad\nu = \frac{d}{2}-1\,,
\ee
where $C_{\ell}^\nu (\xi)$ are the Gegenbauer polynomials as used in \cite{Hogervorst:2013sma}. We can then attempt to solve the superconformal Casimir equations order by order in $s$. Sometimes, the only allowed solution vanishes identically, meaning that the corresponding superconformal multiplet is not allowed to appear in the OPE by the superconformal Ward identities. The following subsection lists the nonzero solutions of the Casimir equation, corresponding to all allowed multiplets.

\subsection{Results}
\label{subsec:susyblocks}

After a careful study of the superconformal Casimir equations in \eqref{eq:sc1} and \eqref{eq:sc2}, we find that the set of multiplets that can appear in our OPE is completely uniform across dimensions (with $2<d\leq 6$). The long multiplets $\mathcal{L}[\Delta,\ell,R]$ are allowed to appear if and only if $R=0$. In addition, there are certain types of short multiplets, namely $\mathcal{B}[\ell,0]$, $\mathcal{B}[\ell,1]$, $\mathcal{D}[0,0]$, $\mathcal{D}[0,1]$, and $\mathcal{D}[0,2]$. Let us now present the superconformal blocks for these multiplets. We recall that $g_{\Delta,l}$ denotes the ordinary conformal block with normalization specified by equations \eqref{confbb}, \eqref{eq:gegenbauer}.

\paragraph{The $\mathcal{L}[\Delta,\ell,0]$ multiplet}
The structure of the $\mathcal{L}[\Delta,\ell,0]$ multiplet leads to the following Ansatz
\begin{equation}\label{Lsusyblock}
\begin{split}
\mathcal{G}_{\Delta,\ell}^{0} &= g_{\Delta,\ell}+ f^0_{2,-2}g_{\Delta+2,\ell-2}+f^{0}_{2,2}g_{\Delta+2,\ell+2}+f^0_{2,0}g_{\Delta+2,\ell}+ f^0_{4,0}g_{\Delta+4,\ell}\;,\\
\mathcal{G}_{\Delta,\ell}^{1} &= f^1_{1,-1}g_{\Delta+1,\ell-1}+f^1_{1,1}g_{\Delta+1,\ell+1} + f^1_{3,-1}g_{\Delta+3,\ell-1}+f^1_{3,1}g_{\Delta+3,\ell+1}\;,\\
\mathcal{G}_{\Delta,\ell}^{2} &= f^2_{2,0}g_{\Delta+2,\ell}\;. \\
\end{split}
\end{equation}
After using the superconformal Casimir equations \eqref{eq:sc1} and \eqref{eq:sc2} and the expansion for conformal blocks discussed around \eqref{confbb}, one can find the explicit form of the coefficient in \eqref{Lsusyblock}.

For $f^{0}_{n,m}$, we find
\begin{align}\label{eq:f0L}
f^0_{2,2}=& \frac{(d+\ell-2) (d+\ell-1) (\Delta +\ell) (\Delta +\ell+2) (\Delta+d +\ell-2)}{4 (d+2 \ell-2) (d+2 \ell) (\Delta +\ell+1) (\Delta +\ell+3) (\Delta-d +\ell+4)}\;,\nonumber\\
f^0_{2,-2}=&\frac{(\ell-1) \ell (\Delta -\ell) (\Delta-d -\ell+4) (\Delta-d -\ell+2)}{4 (d+2 \ell-4) (d+2 \ell-2) (\Delta-d -\ell+5) (\Delta-d -\ell+3) (\Delta-2d -\ell+6)}\;,\nonumber\\
f^0_{2,0}=&\frac{(\Delta +\ell) (\Delta-d -\ell+2)}{6 (d+2 \ell-4) (d+2 \ell)(2 \Delta-d +6) (2 \Delta-d+2)  (\Delta-d +\ell+4) (\Delta-2d -\ell+6)} \times\nonumber\\
& \times\left\{-\Delta ^2 \left[d^2-2 d (4 \ell+5)-8 (\ell-3) (\ell+1)\right]+(d-4) \Delta  \left[d^2-2 d (4 \ell+5)-8 (\ell-3) (\ell+1)\right]+\right.\;\nonumber\\
&\left. +(d-2) \left[-d \ell^2-(d-2) d \ell+2 (d-6) (d-4) (d-3)\right]\right\} \;,\nonumber\\
f^0_{4,0}=& \frac{(\Delta-d +5) (\Delta-d +4) (\Delta-d -\ell+4) (\Delta-d -\ell+2)}{16 (\Delta +\ell+1) (\Delta +\ell+3) (\Delta-d -\ell+5) (\Delta-d -\ell+3) (\Delta-2d -\ell+6)}\times \nonumber \\
& \times \frac{(\Delta +1) (\Delta +2) (\Delta -\ell) (\Delta +\ell) (\Delta +\ell+2) (\Delta+d +\ell-2)}{(2\Delta-d+4) (2\Delta-d+6)^2 (2\Delta-d+8) (\Delta-d +\ell+4)}\;.
\end{align}
For $f^{1}_{n,m}$, the result is
\begin{align}\label{eq:f1L}
f^{1}_{1,-1}=&\frac{\ell (\Delta-d -\ell+2)}{(d+2 \ell-2) (\Delta-2 d -\ell+6)}\;,\nonumber\\
f^{1}_{1,1}=& \frac{(d+\ell-2) (\Delta +\ell)}{(d+2 \ell-2) (\Delta-d +\ell+4)}\;,\nonumber\\
f^{1}_{3,-1}=&\frac{\ell(\Delta +1) (\Delta-d +4) (\Delta -\ell) (\Delta +\ell) }{4  (d+2 \ell-2)(2\Delta -d+4) (2\Delta -d+6) (\Delta-d -\ell+5) }\times\nonumber\\
&\times\frac{(\Delta-d -\ell+4) (\Delta-d -\ell+2)}{(\Delta-d -\ell+3) (\Delta-2d -\ell+6) (\Delta-d +\ell+4)}\;,\nonumber\\
f^{1}_{3,1}=&\frac{(\Delta-d +4) (d+\ell-2) (\Delta-d -\ell+2)}{4 (d+2 \ell-2) (\Delta +\ell+1) (\Delta +\ell+3) (\Delta-2d -\ell+6)}\times\nonumber\\
&\times\frac{(\Delta +1) (\Delta +\ell) (\Delta +\ell+2) (\Delta+d +\ell-2)}{(2 \Delta -d+4) (2\Delta -d+6) (\Delta-d +\ell+4)}\;,
\end{align}
Finally, for $f^{2}_{2,0}$ we have
\begin{align}\label{eq:f2L}
f^{2}_{2,0}=& \frac{(\Delta +\ell) (\Delta-d -\ell+2)}{6 (\Delta-2d -\ell+6) (\Delta-d +\ell+4)}\;.
\end{align}

For unitary SCFTs, the coefficients $f_{n,m}^R$ in the expansion \eqref{ordblocks} of superconformal blocks in terms of the ordinary blocks have to be positive real numbers. This is simply due to the fact that these coefficients are related to the square of certain OPE coefficients. Using the unitarity bounds for the $\mathcal{L}[\Delta,\ell,0]$ multiplet presented in \eqref{5dshort} and \eqref{3dshort}, one can show that indeed all coefficients in \eqref{eq:f0L}, \eqref{eq:f1L}, and \eqref{eq:f2L} are positive real numbers. This constitutes a non-trivial consistency check of our results.  When $d=4$, we can compare our results with the discussion on superconformal blocks in four-dimensional $\mathcal{N}=2$ SCFTs in \cite{Beem:2014zpa}. We find a perfect agreement with the results presented in the Appendix B of their paper.\footnote{In comparing the two sets of results, one should note that our ordinary blocks $g_{\Delta,\ell}$ are related to the ordinary blocks $G_{\Delta}^{(l)}$ of reference \cite{Beem:2014zpa} by $G_{\Delta}^{(l)} = \frac{\ell+1}{2^{\ell}}g_{\Delta,\ell}$.}

In addition to that, we observe that when $d=4$, the coefficients $f_{n,m}^R$ simplify dramatically and one finds the following curious relation between the superconformal blocks and the ordinary non-supersymmetric conformal blocks
\begin{equation}\label{miracle}
\mathcal{G}^{+-+-}_{\Delta,\ell}(z,\bar{z})=\mathcal{G}_{\Delta,\ell}^{0}(z,\bar{z})+\mathcal{G}_{\Delta,\ell}^{1}(z,\bar{z})+\mathcal{G}_{\Delta,\ell}^{2}(z,\bar{z}) = (z\bar{z})^{-1}g_{\Delta+2,\ell} (z,\bar{z})\;.
\end{equation}
This type of relation between superconformal blocks and non-supersymmetric conformal blocks with shifted arguments exists also for SCFTs with four supercharges for any value of $d$, as pointed out in \cite{Bobev:2015jxa}. For theories with eight supercharges, the relation \eqref{miracle} holds only for $d=4$. It will be curious to understand better the reason for the existence of this type of relations.

\paragraph{$\mathcal{B}[\ell,R]$ multiplets}

Due to $SU(2)_R$ selection rules and the superconformal Casimir equations \eqref{eq:sc1} and \eqref{eq:sc2}, the $\mathcal{B}[\ell,R]$ multiplets can appear in the superconformal block expansion only for $R=0$ and $R=1$.

For the type $\mathcal{B}[\ell,0]$ short multiplet, one has $\Delta = l+d-2$ and the following Ansatz for the superconformal blocks
\begin{equation}\label{Bl0susyblock}
\begin{split}
\mathcal{G}_{\ell}^{0} &= g_{l+d-2,\ell}+f^{0}_{2,2}g_{l+d,\ell+2}\;,\\
\mathcal{G}_{\ell}^{1} &= f^{1}_{1,1}g_{l+d-1,\ell+1}\;,\\
\mathcal{G}_{\ell}^{2} &=0\;. \\
\end{split}
\end{equation}
The superconformal Casimir equations determine uniquely the coefficients above
\begin{align}
f^{0}_{2,2}= \frac{(d+\ell-2)^2 (d+\ell-1)}{4 (\ell+1) (d+2 \ell-1) (d+2 \ell+1)}\,,\quad f^{1}_{1,1}= \frac{d+\ell-2}{2 (\ell+1)}\;.
\end{align}

For the $\mathcal{B}[\ell,1]$ multiplet, one has $\Delta=\ell+2d-4$ and the following Ansatz for the superconformal blocks
\begin{equation}\label{Bl1susyblock}
\begin{split}
\mathcal{G}_{\ell}^{0} &= f^{0}_{1,1}g_{\ell+2d-3,\ell+1}+f^{0}_{1,-1}g_{\ell+2d-3,\ell-1}+f^{0}_{3,1}g_{\ell+2d-1,\ell+1}\;,\\
\mathcal{G}_{\ell}^{1} &= g_{\ell+2d-4,\ell}+f^{1}_{2,0}g_{\ell+2d-2,\ell}+f^{1}_{2,2}g_{\ell+2d-2,\ell+2}\;,\\
\mathcal{G}_{\ell}^{2} &= f^{2}_{1,1}g_{\ell+2d-3,\ell+1}\;. \\
\end{split}
\end{equation}
As is familiar by now, the superconformal Casimir equations determines all coefficients in \eqref{Bl1susyblock}
\begin{align}
f^{0}_{1,1}=&~ \frac{2(d+\ell-2) (2 d+\ell-4)}{3(d+2\ell-2) (3 d+2 \ell-4)}\;,\nonumber\\
f^{0}_{1,-1}=&~\frac{(d-2) \ell}{2 (d-1) (d+2 \ell-2)}\;,\nonumber\\
f^{0}_{3,1}=&~\frac{(d-2) (d+\ell-2) (d+\ell-1)^2 (d+\ell) (2 d+\ell-4) (2 d+\ell-3)}{2 (d-1) (\ell+1) (2 d+2 \ell-3) (2 d+2 \ell-1) (3 d+2 \ell-4)^2 (3 d+2 \ell-2)}\;,\nonumber\\
f^{1}_{2,2}=&~\frac{(d+\ell-2) (d+\ell-1)^3 (2 d+\ell-4)}{(\ell+1) (d+2 \ell) (2 d+2 \ell-3) (2 d+2 \ell-1) (3 d+2 \ell-4)}\;,\nonumber\\
f^{1}_{2,0}=&~\frac{(d-2) (d+\ell-2) (d+\ell-1) (2 d+\ell-4)}{(d-1) (d+2 \ell) (3 d+2 \ell-6) (3 d+2 \ell-4)}\;,\nonumber\\
f^{2}_{1,1}=&~\frac{(d+\ell-2)}{3(\ell+1)}\;.
\end{align}
%

\paragraph{$\mathcal{D}[\ell,R]$ multiplets}

Due to the $SU(2)_R$ selection rules and the superconformal Casimir equations \eqref{eq:sc1} and \eqref{eq:sc2}, the $\mathcal{D}[\ell,R]$ multiplets can appear in the superconformal block expansion only for $R=0$, $R=1$ and $R=2$. The $R=0$ multiplet contains only the identity operator.

For the type $\mathcal{D}[0,1]$ multiplet, we have $\Delta=d-2$ and the following Ansatz for the superconformal blocks
\begin{equation}\label{D01susyblock}
\begin{split}
\mathcal{G}^{0} &= f^{0}_{1,1}g_{d-1,1}\;,\\
\mathcal{G}^{1} &= g_{d-2,0}\;,\\
\mathcal{G}^{2} &=0\;. \\
\end{split}
\end{equation}
The superconformal Casimir equations then fix
\begin{align}
f^{0}_{1,1}= \frac{d-2}{2 (d-1)}\;.
\end{align}

For the type $\mathcal{D}[0,2]$ multiplet, we have $\Delta=2d-4$ and the following Ansatz for the superconformal blocks  
\begin{equation}\label{D02susyblock}
\begin{split}
\mathcal{G}^{0} &= f^{0}_{2,0}g_{2d-2,0}\;,\\
\mathcal{G}^{1} &= f^{1}_{1,1}g_{2d-3,1}\;,\\
\mathcal{G}^{2} &=g_{2d-4,0}\;. \\
\end{split}
\end{equation}
The superconformal Casimir equations determine uniquely the coefficients above
\begin{align}
f^{0}_{2,0}= \frac{(d-2)^2}{3(2d-3)(3d-4)}\;, \qquad f^{1}_{1,1}= \frac{d-2}{2d-3}\;.
\end{align}

This completes the derivation of the superconformal blocks for the four-point function of moment map operators. Since the derivation and the final result are quite lengthy, it is important to perform some consistency checks. When $d=4$, superconformal blocks for long and short multiplets in four-dimensional $\mathcal{N}=2$ SCFTs were presented explicitly in \cite{Beem:2014zpa}. Our results above agree with the ones in \cite{Beem:2014zpa} upon setting $d=4$. Another consistency check can be made by considering the moment map operators in the theory of a free hypermultiplet. In Appendix \ref{sec:Freehyper}, we show explicitly how to decompose the four-point function of moment map operators in this theory in terms of our superconformal blocks for any value of $d$.

\section{Discussion}
\label{sec:discussion}

The main focus of our work has been the explicit construction of the superconformal blocks for external moment map operators in SCFTs with eight supercharges. To this end, we have adopted a procedure similar to the one in \cite{Bobev:2015jxa} to treat (at least formally) superconformal algebras and the superconformal quadratic Casimir operator in continuous dimensions $2<d\leq 6$. There are many interesting topics for further studies that could build upon our results.

First, it is clear that the general method for constructing superconformal blocks outlined in this work should be applicable to other external scalar operators, most directly to superconformal primaries of the $\mathcal{D}[0,R]$ multiplet with $R>1$. One of the most important open problems in the theory of superconformal blocks is the construction of the latter when the external operators are the superconformal primaries of the multiplet containing the stress tensor, namely $\mathcal{B}[0,0]$. Substantial progress on this question was made in \cite{Liendo:2015ofa,Ramirez:2016lyk}, but a full formula for the superconformal blocks is still missing. We hope that the superconformal Casimir operator that we derive in this note will prove useful for this problem.

Another interesting extension is to consider external unprotected scalar operators. It has been recently pointed out that one can also make use of the cubic Casimir operator of the superconformal algebra, in addition to the quadratic Casimir operator used in our approach, to derive conformal blocks for external non-protected operators \cite{Cornagliotto:2017dup}. It will be interesting to explore this method for SCFTs with eight supercharges. The construction of superconformal blocks for external operators of non-vanishing spin can also be addressed, although we expect that the results will be significantly more involved.

It is also intriguing to understand better the structure of our superconformal blocks. Recently, it was emphasized that there is a connection between conformal and superconformal blocks and integrability \cite{Isachenkov:2016gim,Schomerus:2016epl,Chen:2016bxc}. This relation has not been explored for superconformal blocks with eight supercharges and our results may shed some light on this story. One particular curiosity that emerged from our calculations is that in $d=4$, we can write the superconformal blocks of long multiplets as ordinary non-supersymmetric conformal blocks with shifted arguments, see \eqref{miracle}. This is reminiscent of the similar situation for SCFTs with four supercharges where for any value of $d\leq4$, one can write the superconformal blocks in terms of shifted non-supersymmetric blocks \cite{Bobev:2015jxa}. It will be interesting to understand the reasons behind this phenomenon and why this structure fails for SCFTs with eight supercharges in $d\neq 4$. 

The results of this paper set the stage for a numerical exploration of the space of SCFTs with eight supercharges in various dimensions. It will be certainly desirable to study the constraints on such theories imposed by unitarity and crossing symmetry using numerical bootstrap methods. This has been quite successful for four-dimensional $\mathcal{N}=2$ \cite{Beem:2014zpa,Lemos:2015awa,Lemos:2015orc} as well as three-dimensional $\mathcal{N}=4$ SCFTs \cite{Chester:2014mea}. A particular fruitful avenue for further progress should be the study of theories in five and six dimensions with exceptional flavor symmetry groups since these arise naturally in string and M-theory.\footnote{This strategy was implemented recently in \cite{Chang:2017xmr} for six-dimensional $(1,0)$ SCFTs.} The advantage offered by our results is that one can perform the numerical analysis for any value of the spacetime dimension $d$. This has proven instructive in the analysis of SCFTs with four supercharges via numerical bootstrap methods \cite{Bobev:2015jxa,Bobev:2015vsa}.

A beautiful algebraic structure spanned by some of the protected operators in four-dimensional $\mathcal{N}=2$, six-dimensional $\mathcal{N}=(2,0)$ and three-dimensional $\mathcal{N}=4$ SCFTs  was uncovered in \cite{Beem:2013sza}. An important open question is whether there is a generalization of this structure for five-dimensional $\mathcal{N}=1$ and six-dimensional $\mathcal{N}=(1,0)$ SCFTs. We hope that the explicit results for short and long superconformal blocks presented in this work will shed some light on this problem.

\bigskip

\noindent \textbf{ Acknowledgements }
\medskip

\noindent We would like to thank Marco Baggio, Chris Beem, Sheer El-Showk, Davide Gaiotto, Fri\dh rik Gautason, Pedro Liendo, Gabriele Tartaglino-Mazzucchelli, Marco Meineri, Miguel Paulos, Silviu Pufu, Emilio Trevisani, and Balt van Rees for useful discussions. The work of NB and EL is supported in part by the starting grant BOF/STG/14/032 from KU Leuven, by an Odysseus grant G0F9516N from the FWO, by the KU Leuven C1 grant ZKD1118 C16/16/005, by the Belgian Federal Science Policy Office through the Inter-University Attraction Pole P7/37, and by the COST Action MP1210 The String Theory Universe. EL is additionally supported by the European Research Council grant no. ERC-2013-CoG 616732 HoloQosmos, as well as the FWO  Odysseus  grants G.001.12 and G.0.E52.14N. The research of DM was supported by Perimeter Institute for Theoretical Physics. Research at Perimeter Institute is supported by the Government of Canada through Industry Canada and by the Province of Ontario through the Ministry of Research and Innovation. DM is grateful to the ITF at KU Leuven for hospitality during various stages of the long gestation period of this project. EL and DM are grateful for the hospitality of the ICTP-SAIFR S\~{a}o Paulo during the completion of this work.

\begin{appendices}

\appendix
\section{Free hypermultiplet check}
\label{sec:Freehyper}
In this appendix, we find the decomposition of the four-point function of the moment map operators in the theory of the free hypermultiplet into our superconformal blocks in general spacetime dimension. The fact that this is possible, and the fact that the resulting coefficients have appropriate positivity properties is a nice consistency check of our formulae for the superconformal blocks.

In the notation of Section \ref{unit-cond}, the hypermultiplet is denoted as $\mathcal{D}[0,1/2]$. Its bottom component consists of two free complex scalars in the doublet of $SU(2)_R$. For our purposes, it is better to think of these as four real scalars $\phi^{p}$, $p=1,\ldots,4$, thus manifesting the full $SO(4)=SU(2)_R\times SU(2)_F$ symmetry group. $SU(2)_R$ is the familiar R-symmetry, while $SU(2)_F$ is a genuine flavor symmetry. We can organize the four real scalars in a $2\times2$ matrix
\be
\phi^{a\dot{a}} = \phi^p \tau_p^{a\dot{a}}\,,
\ee
where $\tau_{1,2,3} = {\rm i} \sigma_{1,2,3}$, $\sigma_p$ being the usual Pauli matrices, and $\tau_4$ the identity matrix. The undotted and dotted indices on $\phi^{a\dot{a}}$ transform as doublets under $SU(2)_R$ and $SU(2)_F$ respectively. The two-point function of $\phi^{a\dot{a}}$ is, up to normalization,
\be
\langle\phi^{a\dot{a}}(x)\phi^{b\dot{b}}(0)\rangle = \frac{\epsilon^{ab}\epsilon^{\dot{a}\dot{b}}}{|x|^{2\nu}}\,,
\ee
where $\nu = (d-2)/2$.

We would like to study the moment map operators for the flavor symmetry $SU(2)_F$, denoted $\varphi^{A\dot{A}}$. The capital undotted and dotted indices transform in the adjoint representation of $SU(2)_R$ and $SU(2)_F$ respectively. Up to normalization, the moment map operators take the form
\be
\varphi^{A\dot{A}} = \sigma^{A}_{ab}\sigma^{\dot{A}}_{\dot{a}\dot{b}}\phi^{a\dot{a}}\phi^{b\dot{b}}\,,
\ee
where $\sigma^{A}_{a b} = \epsilon_{a c}(\sigma^{A})^{c}_{\phantom{c}b}$ with $(\sigma^{A})^{c}_{\phantom{c}b}$ the usual Pauli matrices. The four-point function of $\varphi^{A\dot{A}}$ can be computed using Wick contractions and by the virtue of the $SU(2)_R\times SU(2)_F$ admits the decomposition
\be
\langle \varphi^{A\dot{A}} (x_1) \varphi^{B\dot{B}} (x_2) \varphi^{C\dot{C}} (x_3) \varphi^{D\dot{D}} (x_4) \rangle = 
\frac{1}{(|x_{12}||x_{34}|)^{4\nu}}
\sum\limits_{R,F=0}^{2}Y^{ABCD}_{R}Y^{\dot{A}\dot{B}\dot{C}\dot{D}}_{F}\mathcal{F}^{R F}(u,v)\,,
\ee
with $Y^{ABCD}_{R}$ defined in \eqref{Ys}. Here $F$ stands for the charge under the Cartan of $SU(2)_F$. Since $SU(2)_F$ does not mix with the superconformal symmetry, the functions $\mathcal{F}^{R 0}(u,v)$, $\mathcal{F}^{R 1}(u,v)$, $\mathcal{F}^{R 2}(u,v)$ should each admit a decomposition into our superconformal blocks. We now turn to finding this decomposition for each of these functions.

\subsection{$F=0$ channel}

In the normalization where the identity contributes as 1, the functions $\mathcal{F}^{R 0}(u,v)$ take the following form
\be
\begin{aligned}
\mathcal{F}^{00}(u,v) &=
1+\left(\frac{u}{v}\right)^{\nu  }+u^{\nu }
+\frac{1}{9}\left[\left(\frac{u}{v}\right)^{2 \nu  }+u^{2 \nu  }+\left(\frac{u^2}{v}\right)^{\nu }\right]\,,\\
\mathcal{F}^{10}(u,v) &=
\left(\frac{u}{v}\right)^{\nu  }-u^{\nu }
+\frac{1}{6}\left[\left(\frac{u}{v}\right)^{2 \nu  }-u^{2 \nu  }\right]\,,\\
\mathcal{F}^{20}(u,v) &=
\frac{1}{18}\left[\left(\frac{u}{v}\right)^{2 \nu  }+u^{2 \nu  }-2\left(\frac{u^2}{v}\right)^{\nu }\right]\,.
\end{aligned}
\ee
To find the decomposition of this collection of functions into superconformal blocks means to find an expansion of the following form
\be
\sum\limits_{R=0}^2 Y^{ABCD}_R\mathcal{F}^{R0}(u,v) = \sum\limits_{\mathcal{O}} c_{\varphi\varphi\mathcal{O}}^2 \mathcal{G}^{ABCD}_{\mathcal{O}}(u,v)\,,
\label{eq:sbexpansion}
\ee
with
\be
\mathcal{G}_{\mathcal{O}}^{ABCD}(u,v) = \sum\limits_{R=0}^2 Y^{ABCD}_R\mathcal{G}_{\mathcal{O}}^R(u,v)\,,
\ee
where $\mathcal{G}_{\mathcal{O}}^R(u,v)$ were presented in Section \ref{sec:momentmap}. We find the following expansion
\be
\mathcal{D}[0,0]+\sum\limits_{\ell=0,2,\ldots} \alpha^{(0)}_\ell\mathcal{B}[\ell,0]+
\sum\limits_{\ell=1,3,\ldots} \beta^{(0)}_\ell\mathcal{B}[\ell,1]+
\sum\limits_{\ell,n=0,2,\ldots} \gamma^{(0)}_{n,\ell}\mathcal{L}[4\nu+n,\ell,0]\,,
\ee
where the coefficients $\alpha$, $\beta$, $\gamma$ multiplying each multiplet stand for the $c_{\varphi\varphi\mathcal{O}}^2$ coefficients in the superconformal block expansion \eqref{eq:sbexpansion}. We find the following explicit formulae
\be
\alpha^{(0)}_\ell = \frac{(\nu  )_{\ell/2} (2 \nu  )_{\ell}}{2^{\ell-1}\ell! \left(\nu +\frac{\ell+1}{2}\right)_{\ell/2}}\,,
\ee
\be
\begin{aligned}
\beta^{(0)}_\ell &= \frac{(2 \nu )_\ell (2 \nu ){}^2_{\ell+1} }{3 (\nu )_{\ell+1} (\ell+4 \nu )_{\ell+1}}{\left(\frac{1}{\ell!}-\frac{2^{-2 \ell-1} \left(\frac{\ell+1}{2}+2 \nu \right)_{\frac{\ell+1}{2}}}{\Gamma \left(\frac{\ell+1}{2}\right) \left(\nu +{1}/{2}\right){}^2_{\frac{\ell+1}{2}}}\right)},
\end{aligned}
\ee
\be
\begin{aligned}
\gamma^{(0)}_{n,\ell} &= \frac{(\nu +\ell) \Gamma \left(\frac{n-\ell}{2}+\nu \right)^2 \Gamma (n+4 \nu ) \Gamma \left(\frac{n-\ell}{2}+2 \nu \right) \Gamma \left(\frac{n+\ell}{2}+2 \nu \right)^2\Gamma \left(\frac{n+\ell}{2}+3 \nu \right)}{3 \Gamma (\nu ) \Gamma (2 \nu ) \Gamma \left(\frac{n-\ell}{2}+1\right) \Gamma \left({n}+2\nu -{\ell}\right) \Gamma \left(\frac{n+\ell}{2}+\nu +1\right) \Gamma (3 \nu +n+1)\Gamma \left({n}+{\ell}+4 \nu \right)}\times \,\nonumber \\
&\qquad\qquad \times \left(\frac{(-1)^{\frac{n}{2}+\frac{\ell}{2}} \Gamma \left({n}/{2}+\nu +1\right) \Gamma \left({\ell}/{2}+\nu \right)}{\Gamma (\nu )^2 \Gamma \left({\ell}/{2}+1\right) \Gamma \left({n}/{2}+2 \nu \right)}+\frac{\left(2 \nu +{n}/{2}\right) \Gamma \left( {\ell}+2\nu \right)}{\Gamma (2 \nu )^2 \Gamma (\ell+1)}\right)\, .
\end{aligned}
\ee
In the above and in the following, we use the notation
\be
(a)_{b} \equiv \frac{\Gamma(a+b)}{\Gamma(a)}\,.
\ee
Note that all coefficients are positive, as required by unitarity in our conventions.

\subsection{$F=1$ channel}

In the normalization where the identity contributes as 1, the functions $\mathcal{F}^{R 1}(u,v)$ take the following form
\be
\begin{aligned}
\mathcal{F}^{01}(u,v) &=\left(\frac{u}{v}\right)^{\nu  }-u^{\nu }
+\frac{1}{6}\left[\left(\frac{u}{v}\right)^{2 \nu  }-u^{2 \nu  }\right]\,,\\
\mathcal{F}^{11}(u,v) &=\left(\frac{u}{v}\right)^{\nu  }+u^{\nu }+\frac{1}{4}\left[\left(\frac{u}{v}\right)^{2 \nu  }+u^{2 \nu  }\right]\\
\mathcal{F}^{21}(u,v) &=
\frac{1}{12}\left[\left(\frac{u}{v}\right)^{2 \nu  }-u^{2 \nu  }\right]\,.
\end{aligned}
\ee
This collection of functions can be decomposed into our superconformal blocks with the following result
\be
2\mathcal{D}[0,1]+\sum\limits_{\ell=1,3,\ldots} \alpha^{(1)}_\ell\mathcal{B}[\ell,0]+
\sum\limits_{\ell=0,2,\ldots} \beta^{(1)}_\ell\mathcal{B}[\ell,1]+
\sum\limits_{\ell,n=1,3,\ldots} \gamma^{(1)}_{n,\ell}\mathcal{L}[4\nu+n,\ell,0]\,,
\ee
where
\be
\alpha^{(1)}_\ell = \frac{(\nu  )_{\ell/2} (2 \nu  )_{\ell}}{2^{\ell-1} \ell! \left(\nu +\frac{\ell+1}{2}\right)_{\ell/2}}\,,
\ee
\be
\beta^{(1)}_\ell = \frac{(2 \nu ){}^2_\ell  (2 \nu )_{\ell+1}}{4 \ell! (\nu )_{\ell+1} (\ell+4 \nu )_\ell}.
\ee
\be
\begin{aligned}
\gamma^{(1)}_{n,\ell} &= \frac{(\ell+\nu ) \Gamma (\ell+2 \nu ) \Gamma (n+4 \nu +1) \Gamma \left(-{\ell}/{2}+{n}/{2}+\nu \right)^2 \Gamma \left(-\ell/2+n/2+2 \nu\right) }{4 \Gamma (\ell+1) \Gamma (\nu ) \Gamma (2 \nu )^3 \Gamma \left(-\ell/2+n/2+1\right) \Gamma (n+3 \nu +1) \Gamma (-\ell+n+2 \nu ) }\times \nonumber\\
& \times \frac{\Gamma \left(\ell/2+n/2+2 \nu \right)^2 \Gamma \left (\ell/2+n/2+3 \nu \right)}{\Gamma \left(\ell/2+n/2+ \nu +1\right) \Gamma (\ell+n+4 \nu )}.
\end{aligned}
\ee
Again, all coefficients are positive as they should be.

\subsection{$F=2$ channel}

In the normalization where the identity contributes as 1, the functions $\mathcal{F}^{R 2}(u,v)$ take the following form
\be
\begin{aligned}
\mathcal{F}^{02}(u,v) &=
\frac{1}{18}\left[\left(\frac{u}{v}\right)^{2 \nu  }+u^{2 \nu  }-2\left(\frac{u^2}{v}\right)^{\nu }\right]\,,\\
\mathcal{F}^{12}(u,v) &=\frac{1}{12}\left[\left(\frac{u}{v}\right)^{2 \nu  }-u^{2 \nu  }\right]\,,\\
\mathcal{F}^{22}(u,v) &=\frac{1}{36}\left[\left(\frac{u}{v}\right)^{2 \nu  }+u^{2 \nu  }+4\left(\frac{u^2}{v}\right)^{\nu }\right]\,.
\end{aligned}
\ee
This collection of functions can be decomposed into our superconformal blocks with the following result
\be
\frac{1}{6}\mathcal{D}[0,2]+
\sum\limits_{\ell=1,3,\ldots} \beta^{(2)}_\ell\mathcal{B}[\ell,1]+
\sum\limits_{\ell,n=0,2,\ldots} \gamma^{(2)}_{n,\ell}\mathcal{L}[4\nu+n,\ell,0]\,,
\ee
where
\be
\begin{aligned}
\beta^{(2)}_\ell &= \frac{(2 \nu )_\ell (2 \nu )^2_{\ell+1}}{6 (\nu )_{\ell+1} (\ell+4 \nu )_{\ell+1}} \left(\frac{1}{\ell!}+\frac{4^{-\ell} \left(\frac{\ell+1}{2}+2 \nu \right)_{\frac{\ell+1}{2}}}{\Gamma \left(\frac{\ell+1}{2}\right)\left(\nu +{1}/{2}\right)^2_{\frac{\ell+1}{2}}} \right) 
\end{aligned}
\ee
\be
\begin{aligned}
\gamma^{(2)}_{n,\ell} &= \frac{(\ell+\nu )\Gamma \left(\frac{n+\ell}{2}+2 \nu \right)^2 \Gamma \left(\frac{n-\ell}{2}+\nu \right)^2 \Gamma (n+4 \nu ) \Gamma \left(\frac{n-\ell}{2}+2 \nu \right)\Gamma \left(\frac{n+\ell}{2}+3 \nu \right) }{3 \Gamma (\nu ) \Gamma (2 \nu ) \Gamma \left(\frac{n-\ell}{2}+1\right)\Gamma \left(\nu +\frac{n+\ell}{2}+1\right)\Gamma (3 \nu +n+1)\Gamma(n-\ell+2\nu)\Gamma(n+\ell+4\nu)} \times \nonumber\\
& \qquad\qquad\times \left(\frac{\left(2 \nu +\frac{n}{2}\right) \Gamma \left({\ell}+2\nu \right)}{2 \Gamma (\ell+1) \Gamma (2 \nu )^2}+\frac{(-1)^{\frac{\ell}{2}+\frac{n}{2}+1} \Gamma \left(\frac{\ell}{2}+\nu \right) \Gamma \left(\frac{n}{2}+\nu +1\right)}{\Gamma \left(\frac{\ell}{2}+1\right) \Gamma (\nu )^2 \Gamma \left(\frac{n}{2}+2 \nu \right)} \right)
\end{aligned}
\ee
Once again, all coefficients are positive as they should be.

\end{appendices}

\bibliography{8susyBib}
\bibliographystyle{JHEP}

\end{document}